\documentclass[twocolumn,useAMS,usenatbib]{mn2e}

\usepackage{graphicx}
\usepackage{subfigure}
\usepackage{xcolor}
\usepackage{mathtext,bbm,amsmath,amsfonts,amssymb,indentfirst,syntonly,graphicx}
\usepackage{mathtools}
\usepackage{slashbox}
\usepackage[english]{babel}
\usepackage{calc}
\usepackage{tikz}
\usepackage[T1]{fontenc}
\usepackage{ae,aecompl}

\def\bc{\begin{center}}
\def\ec{\end{center}}
\def\be{\begin{eqnarray}}
\def\ee{\end{eqnarray}}

\title[Testing the DDR using SNe Ia and RSs]{Testing the distance duality relation using type-Ia supernovae and ultra-compact radio sources}
\author[X. Li and H.-N. Lin]
{Xin Li$^{1}$\thanks{e-mail: lixin1981@cqu.edu.cn.} and Hai-Nan Lin$^{1}$\thanks{e-mail: linhn@ihep.ac.cn.}\\
$^{1}$Department of Physics, Chongqing University, Chongqing 401331, China\\}
\begin{document}

\date{Accepted xxxx; Received xxxx; in original form xxxx}

\pagerange{\pageref{firstpage}--\pageref{lastpage}} \pubyear{2017}

\maketitle

\label{firstpage}

\begin{abstract}
  We test the possible deviation of the cosmic distance duality relation $D_A(z)(1+z)^2/D_L(z)\equiv 1$ using the standard candles/rulers in a fully model-independent manner. Type-Ia supernovae are used as the standard candles to derive the luminosity distance $D_L(z)$, and ultra-compact radio sources are used as the standard rulers to obtain the angular diameter distance $D_A(z)$. We write the deviation of distance duality relation as $D_A(z)(1+z)^2/D_L(z)=\eta(z)$. Specifically, we use two parameterizations of $\eta(z)$, i.e. $\eta_1(z)=1+\eta_0 z$ and $\eta_2(z)=1+\eta_0 z/(1+z)$. The parameter $\eta_0$ is obtained using the Markov chain Monte Carlo methods by comparing $D_L(z)$ and $D_A(z)$ at the same redshift. The best-fitting results are $\eta_0=-0.06\pm 0.05$ and $-0.18\pm 0.16$ for the first and second parameterizations, respectively. Our results depend on neither the cosmological models, nor the matter contents or the curvature of the universe.
\end{abstract}

\begin{keywords}
cosmological parameters \--- distance scale \--- supernovae: general
\end{keywords}

\section{Introduction}

In the standard cosmological model, there is a strict correlation between the luminosity distance $D_L(z)$ and the angular diameter distance $D_A(z)$ at the same redshift, i.e. $D_A(z)(1+z)^2/D_L(z)=1$ \citep{Etherington:1933,Etherington:2007}. This is the so-called distance duality relation (DDR). The DDR holds true in any metric theory of gravity such as general relativity, as long as the photons travel along null geodesics and the photon number is conserved. Any deviation of DDR implies that there are new physics beyond the standard cosmological model. Therefore, testing the validity of DDR arouses great interests in recent years.

Many work have been devoted to testing the validity of DDR \citep{Bassett:2003vu,Uzan:2004my,Bernardis:2006,Holanda:2010vb,Piorkowska:2011nhd,Fu:2011,Yang:2013coa,Costa:2015lja,Lv:2016mmq,Holanda:2016msr,Ma:2016bjt,Liao:2016uzb,Chen:2016,Holanda:2016zpz}. All the methods require the measurement of both luminosity distance and angular diameter distance at the same redshift. The luminosity distance $D_L$ can be obtained from type-Ia supernovae (SNe) with high precision. Type-Ia SNe have an approximately consistent absolute luminosity after correcting for stretch and color, and therefore are widely regarded as the standard candles in cosmology \citep{Riess:1998mnb,Perlmutter:1998np}. However, the measurement of angular diameter distance $D_A$ is not as straightforward as that of $D_L$. One way to determine $D_A$ is using the Sunyaev--Zeldovich effect combined with the $X$-ray data from galaxy clusters \citep{Filippis:2005,Bonamente:2006ct}. The obtained $D_A$ in this way depends on the mass model of galaxy clusters, which arouses large uncertainties. Many work have used the $D_L$ data from SNe and $D_A$ data from galaxy clusters to test the DDR, see e.g., \citet{Yang:2013coa} and the references therein. Due to the large uncertainty, most work found no evidence for the violation of DDR. A more accurate measurement of $D_A$ can be obtained from the baryon acoustic oscillations (BAO) \citep{Beutler:2011hx,Anderson:2013zyy,Kazin:2014qga,Delubac:2014aqe}. However, the measurement of BAO requires the statistics of a large number of galaxies, and the number of measured BAO data points so far is very limited. \citet{Ma:2016bjt} used the $D_L$ from SNe and $D_A$ from BAO to test DDR and found 5\% constraints in favor of its validity.

Recently, \citet{Liao:2016uzb} proposed to test the DDR using the angular diameter distance from strong gravitational lensing systems, in combination with the luminosity distance from type-Ia SNe. The angular diameter distance can be deduced from the Einstein radius, and the redshifts of lens and source. However, the gravitation lensing systems could only provide the information of distance ratio between lens and source, and between observer and source, i.e. $R_A\equiv D_{A,ls}/D_{A,s}$. To obtain the distance from observer to lens, a flat FLRW cosmology was assumed, which makes the test of DDR not completely model independent. What's more, the Einstein radius depends on the mass profile of lens, which causes some uncertainty. The luminosity distance is from type-Ia SNe, which is independent of the gravitational lensing systems. The problem is that the SNe and gravitational lensing systems are usually located at different redshifts, making the direct comparison between $D_L$ and $D_A$ impossible. To solve this problem, the authors adopted a matching criterion that if the redshift different between the SNe and lens or source is no more than 0.005, them may be regarded as locating at the same redshift. With such a criterion, there are only 60 or so lensing systems which have matched SNe, although the total number of lensing systems is more than one hundred. The redshift of type-Ia SNe is limited to be $\lesssim 1.4$, much smaller than the redshift of strong gravitation lensing systems. \citet{Holanda:2016msr} combined the SNe, strong gravitational lensing systems and gamma-ray bursts to test the DDR at high redshift, and found that the DDR validity is verified within $1.5\sigma$.

The milliarcsecond ultra-compact radio sources (RSs) provide a unique tool to measure the angular diameter distance. \citet{Kellermann:1993mki} studied the angular size \--- redshift relation  ($\theta-z$ relation) of 79 ultra-compact RSs associated with active galaxies and quasars observed by the very-long-baseline interferometry (VLBI), and showed that the $\theta-z$ relation can be naturally explained by geometrical effect (i.e. the further object has the smaller angular size) in the FLRW cosmology. This implies that the linear size of ultra-compact RSs, $d_0$, is approximately constant, free of evolution effects. From then on, much efforts have been done to standardize the ultra-compact RSs as the cosmological rulers \citep{Jackson:1996stj,Jackson:1997ib,Gurvits:1999hs,Gurvits:1994fgs,Jackson:2004jw,Jackson:2006bg,Jackson:2008ak,Jackson:2012jt}. The possible cosmological evolution of $d_0$ with luminosity $L$, redshift $z$, and spectra index $\alpha$ has been investigated \citep{Gurvits:1999hs,Gurvits:1994fgs,Jackson:2004jw}. With appropriate cutoffs on $L$, $z$ and $\alpha$, the cosmological evolution of $d_0$ can be neglected \citep{Gurvits:1994fgs,Gurvits:1999hs}. Assuming that there is no cosmological evolution of $d_0$, \citet{Jackson:2006bg} used a large sample of RSs data to give a very strict constraint on cosmological parameters.

In this paper, we use the angular diameter distance from ultra-compact RSs, and combine with the luminosity distance from type-Ia SNe, to test the validity of DDR. The main advantage of this method is that it is completely cosmological model-independent. The rest of the paper is organized as follows: In section \ref{sec:method}, we introduce the methodology of testing DDR using type-Ia SNe and ultra-compact RSs data in a model-independent way. In section \ref{sec:results}, we introduce the observational data samples that are used in the test of DDR and give the results. Finally, discussions and conclusions are given in section \ref{sec:conclusions}.

\section{Methodology}\label{sec:method}

In this section, we illustrate how to test the DDR using type-Ia SNe and ultra-compact RSs. Following \citet{Holanda:2010vb}, we write the possible violation of the standard DDR as
\begin{equation}\label{eq:modified_DDR}
  \frac{D_A(z)(1+z)^2}{D_L(z)}=\eta(z).
\end{equation}
We use two different parameterizations of $\eta(z)$,
\begin{equation}
  \eta_1(z)=1+\eta_0z,~~~~\eta_2(z)=1+\eta_0\frac{z}{1+z},
\end{equation}
where $\eta_0$ is a free parameter representing the amplitude of DDR violation. There is no violation of DDR if $\eta_0=0$. By comparing $D_L(z)$ and $D_A(z)$ at the same redshift $z$, we can constrain the parameter $\eta_0$.

The luminosity distance can be derived from type-Ia SNe. Type-Ia SNe are widely used as the standard candles in cosmology due to their approximately consistent absolute luminosity \citep{Riess:1998mnb,Perlmutter:1998np}. The distance modulus of SNe can be extracted from the light curves using the following empirical relation \citep{Tripp:1998,Guy:2005,Guy:2007,Suzuki:2012dhd,Betoule:2014frx}
\begin{equation}\label{eq:mu_sn}
  \mu_{\rm sn}=m_B^*-M_B+\alpha X_1-\beta \mathcal{C}.
\end{equation}
where $m_B^*$ is the apparent magnitude, $M_B$ is the absolute magnitude, $X_1$ and $\mathcal{C}$ are the stretch factor and color parameter, respectively. The two parameters $\alpha$ and $\beta$ are universal constants and can be fitted simultaneously with cosmological parameters, or more generally, can be marginalized over. The uncertainty of $\mu_{\rm sn}$ is propagated from the uncertainties of $m_B^*$, $X_1$ and $\mathcal{C}$ using the standard error propagation formula,
\begin{equation}
  \sigma_{\mu_{\rm sn}}=\sqrt{\sigma_{m_B^*}^2+\alpha^2\sigma_{X_1}^2+\beta^2\sigma_{\mathcal{C}}^2}.
\end{equation}

The angular diameter distance is derived from the ultra-compact RSs. Due to the approximately constant linear size of RSs, the angular diameter distance can be easily obtained if the angular size $\theta$ is observed,
\begin{equation}\label{eq:DA}
  D_{A,\rm rs}=\frac{d_0}{\theta},
\end{equation}
where the linear size $d_0$ is not known a prior and is regarded as a free parameter. Combining Eqs.(\ref{eq:modified_DDR}) and (\ref{eq:DA}), the luminosity distance of RSs can be written as
\begin{equation}\label{eq:DL_rs}
  D_{L,\rm rs}=\frac{d_0}{\theta}\frac{(1+z)^2}{\eta(z)}.
\end{equation}
The distance modulus of RSs is given by
\begin{equation}\label{eq:mu_rs}
  \mu_{\rm rs}=5\log_{10}\frac{D_{L,{\rm rs}}}{\rm Mpc}+25.
\end{equation}
The uncertainty of $\mu_{\rm rs}$ is propagated from that of $\theta$,
\begin{equation}\label{eq:mu_rs_error}
  \sigma_{\mu_{\rm rs}}=\frac{5}{\ln 10}\frac{\sigma_{\theta}}{\theta},
\end{equation}
where `\,ln' is the natural logarithm.

Difficulties arise when we try to directly compare $\mu_{\rm sn}$ with $\mu_{\rm rs}$. This is because SNe and RSs usually locate at different redshift. For a specific SN, there is in general no RS locating at the same redshift, and vice visa. To solve this problem, we first reconstruct the $\theta-z$ relation using the Gaussian processes \citep{Seikel:2012uu}. Then the $\mu_{\rm rs}(z)$ function can be reconstructed using Eqs.(\ref{eq:DL_rs}) -- (\ref{eq:mu_rs_error}). Given $\mu_{\rm rs}(z)$ function, we can calculate the distance modulus of RSs at any desired redshift. We use the reconstructed $\mu_{\rm rs}(z)$ function to calculate the distance modulus of RS at the redshift of each SNe. Therefore, the comparison between $\mu_{\rm rs}$ and $\mu_{\rm sn}$ at the same redshift is possible.

We use the Markov chain Monte Carlo methods \citep{ForemanMackey:2012ig} to calculate the posterior probability distribution functions of free parameters. The likelihood is given by
\begin{equation}\label{eq:likelihood}
  \mathcal{L(\mathbf p)}=\prod\frac{1}{\sqrt{2\pi}\sigma_\mu}\exp\left[-\frac{1}{2}\left(\frac{\mu_{\rm sn}-\mu_{\rm rs}}{\sigma_\mu}\right)^2\right],
\end{equation}
where
\begin{equation}\label{eq:dmu_total}
  \sigma_\mu=(\sigma_{\mu_{\rm sn}}^2+\sigma_{\mu_{\rm rs}}^2+\sigma_{\rm int}^2)^{1/2}
\end{equation}
is the uncertainty of distance modulus of SNe and RSs, $\mathbf{p}=(d_0,\eta_0,\alpha,\beta, M_B, \sigma_{\rm int})$ is the set of free parameters, and the product runs over all the SN-RS pairs. We have added the intrinsic scatter term in Eq.(\ref{eq:dmu_total}) to account for any other uncertainties. Note that $d_0$ is degenerated with $M_B$, so they cannot be constrained simultaneously. One parameter should be fixed in order to constrain the other. We leave $d_0$ free and fix $M_B=-19.32$ \citep{Suzuki:2012dhd}.

\section{Data and Results}\label{sec:results}

The SNe sample is taken from \citet{Suzuki:2012dhd}, i.e. the Union2.1 sample. Union 2.1 consists of 580 type-Ia SNe with high light curve quality in the redshift range $[0.015,1.414]$. Each SN has well measured redshift $z$, apparent magnitude $m_B^*$, stretch factor $X_1$ and color factor $\mathcal{C}$. Note that \citet{Suzuki:2012dhd} also published the distance modulus of each SNe, which is calibrated in the $w$CDM model. We use the original light curve parameters rather than the published distance moduli in order to avoid the model dependence. The distance moduli are calculated using Eq.(\ref{eq:mu_sn}), with $\alpha$ and $\beta$ as free parameters.

\begin{figure}
\centering
  \includegraphics[width=0.5\textwidth]{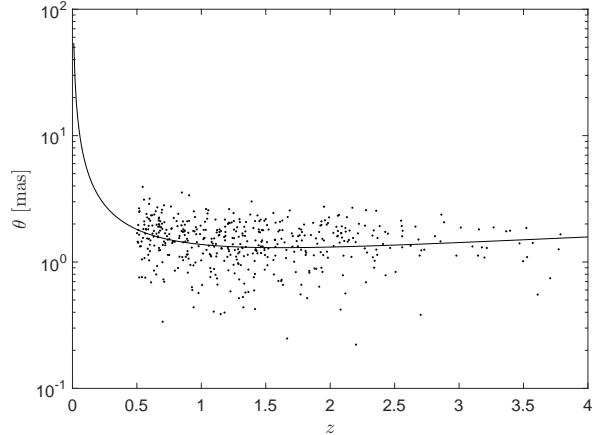}
  \caption{\small{The 468 radio sources in the $\theta-z$ plane. The black line is the best-fitting results to $\Lambda$CDM model, with the best-fitting parameters $\Omega_M=0.29\pm 0.10$, $d_0=7.76\pm 0.57~h_0^{-1}$ pc.}} \label{fig:logTheta}
\end{figure}

The RSs sample is taken from \citet{Jackson:2006bg}. The sample is selected from a complication of VLBI survey of ultra-compact RSs at 2.29 GHz released by \citet{Preston:1985bgd}. The redshift and radio flux are updated according to the recent observations. The sample consists of 613 RSs in the redshift range $[0.0035,3.787]$. Among the sample, there are 468 RSs which have redshift larger than 0.5. Data with $z<0.5$ are not appropriate to use as standard rulers because they are more affected by cosmological evolutions \citep{Gurvits:1994fgs,Jackson:2004jw,Jackson:2006bg,Jackson:2008ak,Jackson:2012jt}. The 468 RSs data in the $\theta-z$ plane are plotted in Figure \ref{fig:logTheta}.
To show that the RSs data can be used to constrain the cosmological parameters, we fit the 468 RSs data points to the standard $\Lambda$CDM model. We get the best-fitting parameters $\Omega_M=0.29\pm 0.10$, $d_0=7.76\pm 0.57~h_0^{-1}$ pc. Here $h_0$ is the Hubble constant in unit of $100~{\rm km}~{\rm s}^{-1}~{\rm Mpc}^{-1}$. Although has a large uncertainty, the $\Omega_M$ value is well consistent with the much more precise result from CMB observations \citep{Ade:2015xua}.

\begin{figure}
\centering
  \includegraphics[width=0.5\textwidth]{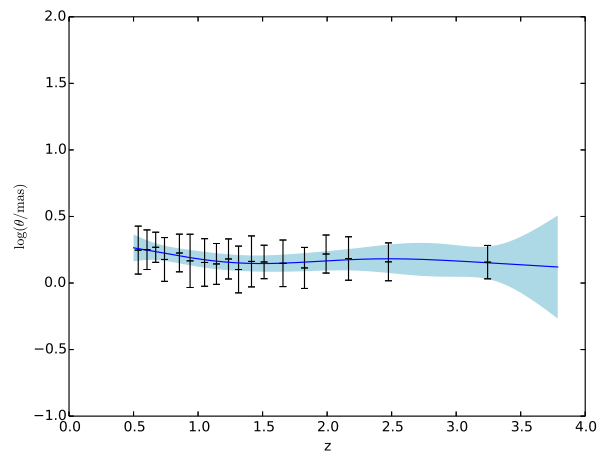}
  \caption{\small{The $\theta-z$ relation in the redshift range $0.5<z<3.787$ reconstructed from the Gaussian processes. The blue region is the $1\sigma$ uncertainty.}} \label{fig:GP_Theta_z}
\end{figure}

From Figure \ref{fig:logTheta}, we can see that the original data points show large scatter around the $\theta-z$ theoretical curve. Following Refs. \citep{Kellermann:1993mki,Jackson:1996stj,Jackson:1997ib,Gurvits:1999hs,Gurvits:1994fgs,Jackson:2004jw,Jackson:2006bg}, we bin the raw data into 18 bins, with 26 data points in each bin. The central value and $1\sigma$ uncertainty are taken to be the mean and standard deviation of data points in each bin, respectively. The binned data are plotted in Figure \ref{fig:GP_Theta_z}.
Based on the binned data, we use the publicly available python package \textsf{GaPP} \citep{Seikel:2012uu} to reconstruct the $\theta(z)$ function in the redshift range $[0.5, 3.787]$. The Gaussian processes only depends on the covariance function. Here we adopt the most widely used squared exponential covariance function. The results are plotted in Figure \ref{fig:GP_Theta_z}. The blue range stands for the $1\sigma$ uncertainty. In the reconstruction, we use $\log\theta$ instead of $\theta$. This is because in the likelihood function Eq.(\ref{eq:likelihood}), we use the distance modulus rather than the luminosity distance, while the former is linearly correlated with $\log\theta$. In principle, the Gaussian processes can reconstruct function at any point. Beyond the data range, however, the reconstructed function has too large uncertainty. The reconstructed $\theta(z)$ function is used to calculate the distance modulus of RSs at the redshift of SNe according to the method described in the last section. SNe with $z<0.5$ are ignored. There are 167 SNe remain in the redshift range $0.5<z\leqslant 1.414$.

\begin{figure}
\centering
  \includegraphics[width=0.5\textwidth]{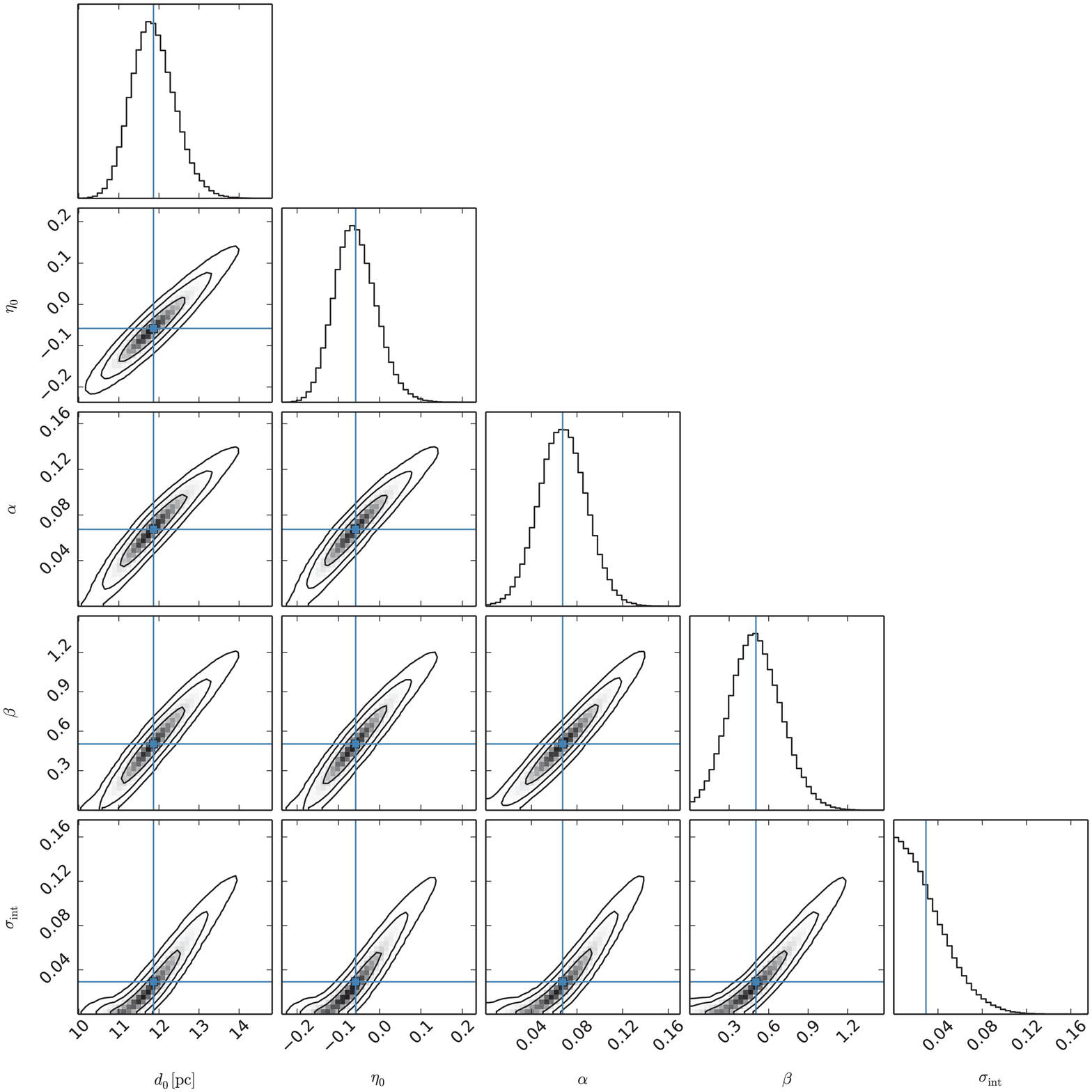}
  \caption{\small{The marginalized likelihood distributions and 2-dimensional confidence regions for the parameters $d_0$, $\eta_0$, $\alpha$, $\beta$ and $\sigma_{\rm int}$ in the first parametrization $\eta_1(z)=1+\eta_0 z$. The blue lines represent the mean values.}}\label{fig:triangle}
\end{figure}

\begin{figure}
\centering
  \includegraphics[width=0.5\textwidth]{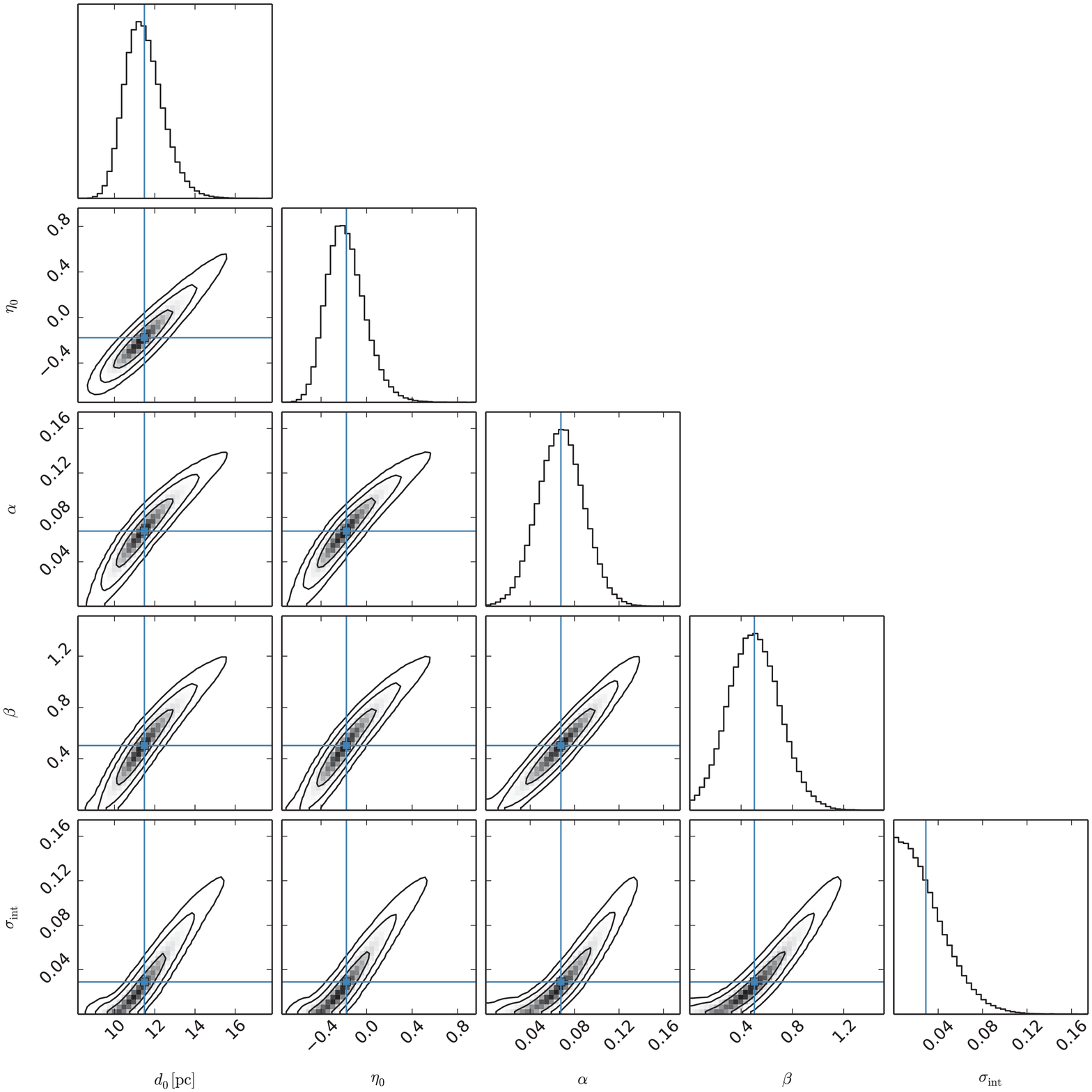}
  \caption{\small{The marginalized likelihood distributions and 2-dimensional confidence regions for the parameters $d_0$, $\eta_0$, $\alpha$, $\beta$ and $\sigma_{\rm int}$ in the second parametrization $\eta_2(z)=1+\eta_0 z/(1+z)$. The blue lines represent the mean values.}}\label{fig:triangle2}
\end{figure}

We use the publicly available python package \textsf{emcee} \citep{ForemanMackey:2012ig} to do the Markov chain Monte Carlo analysis. The likelihood function is given by Eq.(\ref{eq:likelihood}). We use a flat prior on the parameters: $P(d_0)=U[1,20]$ pc, $P(\eta_0)=U[-1,1]$, $P(\alpha)=U[0,1]$, $P(\beta)=U[0,10]$, and $P(\sigma_{\rm int})=U[0,1]$. The absolute magnitude $M_B$ is fixed to $-19.32$ due to the degeneracy between $d_0$ and $M_B$. The marginalized likelihood distributions and the 2-dimensional confidence regions for the parameters are plotted in Figure \ref{fig:triangle} and Figure \ref{fig:triangle2} for the first and second parameterizations, respectively. The best-fitting (mean) values and $1\sigma$ errors (standard deviations) of the parameters are listed in Table \ref{tab:parameter}.
\begin{table*}
\caption{\small{The best-fitting parameters and their $1\sigma$ uncertainties in two different parameterizations of $\eta(z)$. For the intrinsic scatter, only the $1\sigma$ upper limit can be obtained.}}
\begin{tabular}{cccccc}
  \hline\hline
  & $d_0$ [pc] & $\eta_0$ & $\alpha$ & $\beta$ & $\sigma_{\rm int}$\\
  \hline
  $\eta_1(z)$ & $11.86\pm 0.54$ & $-0.06\pm 0.05$ & $0.07\pm 0.02$ & $0.50\pm 0.20$ & $<0.05$ \\
  $\eta_2(z)$ & $11.46\pm 0.92$ & $-0.18\pm 0.16$ & $0.07\pm 0.02$ & $0.50\pm 0.20$ & $<0.05$ \\
  \hline
\end{tabular}\label{tab:parameter}
\end{table*}
Except for the intrinsic scatter $\sigma_{\rm int}$, all the other parameters can be well constrained. The $1\sigma$ upper limit of $\sigma_{\rm int}$ is $0.05$, which is much smaller than the errors of SNe and RSs, and therefore can be neglected. In fact, if we fix $\sigma_{\rm int}=0$, all the other parameters are almost unaffected. The best-fitting linear size is $d_0=11.86\pm 0.54$ pc in the first parametrization and $d_0=11.46\pm 0.92$ pc in the second parametrization. These are well consistent with the result of \citet{Jackson:2012jt} obtained in $\Lambda$CDM cosmology, i.e. $7.76~h_0^{-1}$ pc (assuming $h_0=0.67$ \citep{Ade:2013zuv,Ade:2015xua}). The best-fitting $\eta_0$ is $-0.06\pm 0.05$ ($-0.18\pm 0.16$) in the first (second) parametrization. In both parameterizations, there is no strong evidence for the violation of DDR. Note that we make no assumption on the matter contents or the curvature of the universe, and therefore our results are completely model-independent.

\section{Discussions and conclusions}\label{sec:conclusions}

The DDR, which relates the luminosity distance $D_L$ to the angular diameter distance $D_A$ at the same redshift $z$, plays an important role in modern cosmology and astronomy. It is a natural inference of Etherington reciprocity theorem. Any violation of DDR may imply new physics beyond the standard model. Thus, testing the validity of DDR is essential and has aroused great interests in recent years. The most ideal way to test DDR is to measure $D_L$ and $D_A$ to a specific object (hence $D_L$ and $D_A$ are measured at the same redshift, of course). However, at the present day it is difficult to directly measure both $D_L$ and $D_A$ to a celestial object. The most popular way is to measure $D_L$ from type-Ia SNe, and $D_A$ from other objects such as galaxy clusters, BAO, or strong gravitational lensing systems. Although $D_L$ from SNe is free of cosmological model and has high accuracy, $D_A$ from other objects is more or less model-dependent. For example, $D_A$ from galaxy clusters depends on the mass profile of clusters, $D_A$ from BAO depends on the cosmological models, and $D_A$ from strong gravitational lensing systems depends on the curvature of the universe and the mass profile of the lens. On the other hand, the uncertainty of $D_A$ is much larger compared to the uncertainty of $D_L$ from SNe. Due to the large uncertainty, previous tests show no strong evidence for the violation of DDR.

In this paper, we used the type-Ia SNe as the standard candles and ultra-compact RSs as the standard rulers to test the validity of DDR. The linear size of ultra-compact RSs was assumed to be approximately constant, so the angular diameters distance $D_A$ can be obtained from the measured angular size $\theta$. In order to compare $D_A$ from RSs to the luminosity distance $D_A$ from SNe at the same redshift, the Gaussian processes were used to reconstruct $D_A(z)$ function. We parameterized the violation of DDR in two different forms, and used the Markov chain Monte Carlo method to calculate the posterior probability density function of each parameters. The violation amplitudes in the first and second parameterizations are $\eta_0=-0.06\pm 0.05$ and $-0.18\pm 0.16$, respectively. These are consistent with the results of \citet{Liao:2016uzb} obtained from SNe and strong gravitational lensing systems (plus galaxy clusters), but the uncertainty was much reduced compared to the previous results. In both parameterizations, no strong evidence for the violation of DDR was found. The main advantage of our method is that it is completely independent of cosmological model and the curvature of the universe. What's more, it is also independent of the mass profile of galaxies or clusters. In both parameterizations, $\eta_0$ is negative within $1\sigma$ uncertainty. The negative value of $\eta_0$ means that $\eta(z)$ is smaller than unity. This may happen if the photons from SNe are not conserved but some of them are absorbed by the intergalactic medium. This makes SNe seem to be dimmer than expected, which just looks as if the SNe locate at a further distance.

We note that the $\alpha$ and $\beta$ values obtained here are much smaller than that of \citet{Suzuki:2012dhd}. There are many reasons which can cause the discrepancy. First, our results are obtain from the combination of RSs and SNe, while \citet{Suzuki:2012dhd} obtained the $\alpha$ and $\beta$ values from the pure SNe. The uncertainty of RSs is much larger than that of SNe. Second, only SNe with $z>0.5$ are used in our paper, while in \citet{Suzuki:2012dhd} the full Union2.1 sample was used. Finally, biases from the selection effects in both the SNe and RSs samples may be the most important reason. Especially, the SNe sample used here is at high redshift $(z>0.5)$, where the biases are much worse. Dealing with biased samples is beyond the scope of our present work. The different nuisance parameters only cause the discrepancy of SNe distance by 0.1 mag, which is much smaller than the uncertainty of distance of RSs. The main conclusions of our manuscript, therefore, are not affected by the nuisance parameters.

The main shortcoming is that the low redshift ($z<0.5$) RSs couldn't be used as the standard rulers, while most SNe has redshift $z<0.5$. In the redshift overlapping range ($0.5<z<1.414$), there are only 167 SNe, much smaller than the full Union2.1 sample. Nevertheless, this subsample of SNe is already much larger than the available strong gravitational lensing systems or the galaxy clusters. We may enlarge the sample by adding some high-redshift data such as gamma-ray bursts to SNe sample, or adding some low-redshift galaxy cluster to the RSs sample. The price, however, is that the uncertainty is also enlarged.

\section*{Acknowledgements}
This work has been supported by the National Natural Science Fund of China (Grant Nos. 11603005, 11775038 and 11647307), and the Fundamental Research Funds for the Central Universities (Grant No. 106112016CDJCR301206).

\label{lastpage}

\end{document}